\documentclass[showpacs,preprintnumbers,amsmath,amssymb, prd,11pt]{revtex4}

\usepackage{amsmath}
\usepackage{graphicx}
\usepackage{amssymb}
\usepackage{amsfonts}
\usepackage{latexsym}

\def\beq{\begin{eqnarray}}
\def\eeq{\end{eqnarray}}
\def\ln{\,\mbox{ln}\,}

\def\det{\,\mbox{det}\,}

\def\diag{\,\mbox{diag}\,}

\renewcommand{\diag}{\,\mbox{diag}\,}

\def\al{\alpha}
\def\be{\beta}

\def\de{\delta}

\def\ka{\kappa}
\def\la{\lambda}
\def\na{\nabla}
\def\pa{\partial}

\def\si{\sigma}

\def\ph{\varphi}

\def\th{\theta}

\def\Om{\Omega}

\begin{document}

\begin{flushright}
RESCEU-25/10
\end{flushright}
\vskip 0.7cm


\title{Auxiliary fields representation for modified gravity models}


\author{Davi C. Rodrigues}\email{davirodrigues.ufes@gmail.com}
\affiliation{Departamento de F\'isica, CCE, 
Universidade Federal do Esp\'irito Santo, 29075-910, Vit\'oria, ES, Brazil} 
\author{Filipe de O. Salles}\email{salles@ice.ufjf.br}  
\author{Ilya L. Shapiro}\email{shapiro@fisica.ufjf.br . Also at Tomsk State 
Pedagogical University, Tomsk, Russia.}
\affiliation{Departamento de F\'{\i}sica, ICE, 
Universidade Federal de Juiz de Fora, 36036-330, MG, Brazil} 
\author{Alexei A. Starobinsky}\email{alstar@landau.ac.ru}
\affiliation{Landau Institute for Theoretical Physics, Moscow, 119334, Russia
\\ 
RESCEU, Graduate School of Science, The University of Tokyo, Tokyo 113-0033, 
Japan}




\begin{abstract}
We consider tensor-multiscalar representations for several types of 
modified gravity actions. The first example is the theory with the 
action representing an arbitrary smooth function of the scalar 
curvature $R$ and $\Box R$, the integrand of the Gauss-Bonnet term 
and the square of the Weyl tensor. We present a simple procedure 
leading to an equivalent theory of a space-time metric and four 
auxiliary scalars and specially discuss calibration of a cosmological
constant and the condition of the existence of $dS$-like solutions
in the case of empty universe. The condition for obtaining a
smaller number of independent scalar fields is derived. The second
example is the Eddington-like gravity action. In this case we
show, in particular, the equivalence of the theory to GR with the
cosmological constant term, with or without use of the first-order
formalism, and also discuss some possible generalizations.
\end{abstract}



\maketitle

\section{Introduction}

Recently, there was a considerable interest in the $f(R)$ theories
described by the action 
\beq
S_{f}\,\,=\,\,\int\,d^{4}x\,\sqrt{-g}\,f(R)\,, 
\label{1} 
\eeq
where $f(R)$ is some differentiable function, see \cite{faraoni}
for recent reviews. It is well-known that, under the condition
$f''(R)\not= 0$, the theory is dynamically equivalent to
scalar-tensor theory of gravity with the potential depending on
the form of the function $f(R)$. Our purpose is to discuss this
equivalence in a slightly different framework. As an application
of our method we will be able to generalize the equivalence
theorem to the more general case when the action depends on a
function of many variables, $f(X^i)$, with $X^i$ being, e.g., $R$,
$\Box R$, Gauss-Bonnet integrand $E$ and/or other quantities. Let
us note that quantum corrections to GR (coming from the
semiclassical approach to quantum gravity or from the string
theory) can be modelled by such a function to some extent. Another
advantage of the method which we present here is that it can be
used, also, for other theories, e.g. for Eddington-like models.
Last, but not least, our method is a bit more explicit and simple
than the previously known ones, e.g.,
\cite{O_Hanlon,TT,Chiba,Whitt,wands,GSS,SDO}). All considerations
will be presented for the $D \,=\, 4$ case in order to make them
more explicit, but they can be more or less straightforwardly
generalized for an arbitrary $D \neq 2$ case.

In the present paper we will mainly follow Lagrangian approach, 
but it is worthwhile to mention that there are also some recent 
papers treating $f(R)$ \cite{Deru1} and even more general 
$f(R_{\mu\nu\al\be})$ \cite{Deru2} theories within the 
canonical formalism. Earlier, the discussion of stability 
issues for a while class of higher derivative theories has
been presented in \cite{H.J.Sch-94} at both classical level 
and also for the Wheeler-DeWitt equation. In general, the
equivalence of two theories at the classical level does not 
imply their equivalence at the quantum level. The quantum 
treatment of higher derivative and scalar-tensor theories is 
not a trivial issue, hence here we present a short discussion of 
it only. 

The paper is organized as follows. In Sect. 2 we consider the
simplest case of the theory (\ref{1}) and describe a simple way of
mapping it into a metric-scalar (scalar-tensor) model. The content
of this section is mainly not original, we just give a bit more
simple form of the known transformations. One of relatively new
aspects of our consideration is the procedure to fix the
cosmological constant term in the metric-scalar representation of
the theory. We also check this procedure by using the dS-like
exponential solution. In Sect. 3 we address gravity theories of a
rather generic form, in which the action contains an arbitrary
function of various scalar, curvature-dependent invariants, such
as the scalar curvature $R$, the Gauss-Bonnet term, the square of the
Weyl tensor and others. We develop a systematic approach to map
such theories into the metric-multiscalar models. One of the new
elements of our consideration is that it includes the case when
the initial theory has such Hamiltonian constraints that the number
of independent auxiliary scalars is smaller than the number
of initial curvature-dependent invariants. The condition for this
to occur is obtained. Sect. 4 is devoted to the formulation of
general conditions for the existence of exponential solutions
in the theories investigated in Sect. 3. In Sect. 5 we consider the
theory which is based on the string low-energy effective action of
gravity, up to the third order in curvature invariants. In this case
one cannot construct an equivalent scalar-tensor representation at
the level of action or general equations of motion for the metric
field. However, this problem can be perfectly addressed for the
much more restricted case of a homogeneous and isotropic
cosmological solution. In Sect. 6 we apply our method of
constructing equivalent theories to the wide class of the
Eddington-like gravity theories. It is shown how to construct
dual theories for such gravity theories, including both the second
order and Palatini formalisms. In the latter case the auxiliary
tensor field can be interpreted as a space-time metric. In Sect.
7 we discuss the equivalence between different representations of
the theory (\ref{1}) and its generalizations at the quantum level. 
Finally, in Sect. 8 we draw our conclusions.

\section{Equivalence of $f(R)$ and metric-scalar theory}

Let us start from a simple pedagogical example of the theory
(\ref{1}) and find its metric-scalar (scalar-tensor) dual.
Consider the theory described by the action
\beq
S_{1}\,\,=\,\,\int\,d^{4}x\,\sqrt{-g}\, \{\psi\,R - V(\psi)\}\,.
\label{2}
\eeq
The theory (\ref{2}) describes a dynamical scalar
$\psi$, despite there is no kinetic term for $\psi$ in the action.
One can establish the relation between the theories
(\ref{1}) and (\ref{2}). The equation of motion which follows from
the variation of $\psi$ in (\ref{2}) has the form
\beq
R\,=\,V^\prime(\psi)\,=\,\frac{dV}{d\psi}\,.
\label{1.8}
\eeq
After solving (\ref{1.8}) with respect to $\psi$ and substituting
this solution $\psi=\psi(R)$ back into (\ref{2}), we obtain the
action (\ref{1}) with \beq \psi(R)\cdot R \,-\,V
\big(\psi(R)\big)\,=\,f(R)\,. \label{1.9} \eeq This means that the
equivalence of the two actions is dynamical, i.e. it holds on
extremal curves of the field $\psi$. Later on we shall confirm the
validity of this procedure through the equations of motion for
both metric and $\psi$, i.e. in a way similar to the one of
\cite{Whitt}.

Our next step will be to find the relation between the functions
$V(\psi)$ and $f(R)$. Taking the derivative $\,d/dR\,$ of Eq.
(\ref{1.9}), we arrive at the relation
\beq
\psi \,+\,
R\,\psi^\prime(R) \,-\, V^\prime(\psi)\, \psi^\prime(R)
\,=\,f^\prime(R)\,.
\label{2.0}
\eeq
In this formula we assume
that $\psi=\psi(R)$ and $R=R(\psi)$. Using (\ref{1.8}), the
equation (\ref{2.0}) immediately reduces to the very simple
relation \beq \psi\, = \,f^\prime(R), \label{2.1} \eeq indicating
that the function $R\,=\,V^\prime(\psi)$ is nothing else but the
inverse to the function $\psi\,=\,f^\prime(R)$.

Finally, we arrive at the following receipt for deriving
the potential $V(\psi)$ for a given $f(R)$.

a) Calculate $\psi\,=\,f^\prime(R)$ and invert it, obtaining
$R\,=\,V^\prime(\psi)$. Note that the possibility of such
inversion requires $f''(R)\not= 0$.

b) Integrate over $\psi$:
\beq
V(\psi)\,=\,\Omega_0\,+\,\int_0^\psi
R(\psi)\,d\psi\,.
\label{2.2}
\eeq
One has to note that an
arbitrary integration constant $\,\Omega_0\,$ in (\ref{2.2})
exactly corresponds to the constant $\,f_0=f(R=0)\,$ component of
the integrand of Eq. (\ref{1}), which is indeed lost when we take
the derivative $\,f^\prime(R)$. Furthermore, in order to fix the
constant $\Omega_{0}$, one can use the following simple
consideration. By using (\ref{1.9}) we arrive at
\beq
V(\psi)\,=\,
R\,\psi - f(R)\,, \quad \mbox{where} \quad \psi\,=\,f^\prime(R)\,.
\label{most3}
\eeq
Remember that when placed into the covariant
action, $\Omega_{0}$ can not be regarded as an irrelevant
constant, because it is multiplied by the metric-dependent factor
$\sqrt{-g}$. As far as (\ref{2.2}) should be equal to
(\ref{most3}), one can then fix $\Omega_{0}$. Later one we will
additionally check the validity of this procedure for a
cosmological $dS$-like solution.

The prescription given above enables one, in principle, to
find the potential function $V(\psi)$ for a given $f(R)$.
Let us check the results of this simple procedure
at the level of the equations of motion.
Taking variation of the equation (\ref{1}) with respect to
the metric, we obtain
\beq
f^{\prime}\big(R_{\mu\nu} - \frac{1}{2}R g_{\mu\nu}\big)
+ \frac{1}{2} g_{\mu\nu}(R\,f^{\prime} - f)
- \nabla_{\mu}\nabla_{\nu}f^{\prime} + g_{\mu\nu}\Box f \,=\, 0\,.
\label{4}
\eeq
Performing the same operation for (\ref{2}), we arrive at
\beq
\psi \Big(R_{\mu\nu} - \frac{1}{2}R g_{\mu\nu}\Big)\,
\,=\,\, - \frac{1}{2} g_{\mu\nu}V(\psi)
+ \nabla_{\mu}\nabla_{\nu}\psi - g^{\mu\nu}\Box \psi\, \,=\,\, 0\,.
\label{5}
\eeq
One can verify that the equivalence between (\ref{4}) and (\ref{5})
holds if the relation
\beq
\frac{f^{\prime} R - f}{f^{\prime}}\, \,=\,\, \frac{V(\psi)}{\psi}\,
\label{6}
\eeq
is satisfied. It is easy to check that the solution of this equation
has the form (\ref{2.1}).

Consider some particular example for the procedure described
above. The simplest case leading to the linear equations
$\psi=f^\prime(R)$ is \beq f(R) \,\,=\,\, \Omega - \kappa^{2} R +
\frac{\alpha}{2} R^{2}\,. \label{quadratic} \eeq Using our
previous results, one can easily arrive at \beq \psi(R)
\,=\,f^\prime(R)\,=\, - \ka^{2} + \al R \qquad \Longrightarrow
\qquad V^\prime(\psi) \,=\, R \,=\, \frac{\psi + \ka^{2}}{\al}\,.
\label{105} \eeq Integrating (\ref{105}) we get \beq
V(\psi)\,\,\,=\,\,\,\Omega_{0} + \frac{\psi^{2}}{2\alpha} +
\frac{\kappa^{2}\psi}{\alpha}\,. \label{100} \eeq Finally, in
order to fix the integration constant, one has to put
$\psi\,=\,-\ka^{2} + \al\,R$ back into (\ref{100}) and
compare it to (\ref{quadratic}). This procedure gives us \beq
\Om\,=\,\Om_{0} - \frac{\ka^4}{2}\,. \eeq

One can perform a simple verification of the described procedure
for fixing $\Om$. For this end, we will now derive the $dS$-like
solution for both theories (\ref{1}) and (\ref{2}) in case of
(\ref{quadratic}). The metric of our interest is
$$
ds^2 = g_{\mu\nu}dx^\mu dx^\nu
= dt^2 - a^2(t)\,\Big(\frac{1}{1-kr^2} + r^2 \,d\Om \Big)\,,
$$
$a(t)=\exp \{ \si(t)\}$ and afterwards we will set \ $\si(t)=H_0t$.
It is easy to obtain the equation for $\si$ for the theory (\ref{2}),
\beq
\frac{1}{\sqrt{-g}}\,\frac{\delta S_{1}}{\delta \sigma}
\,=\, -6\,e^{-2\sigma}(2\psi\,k + \psi^{\prime\prime}
+ 2\sigma^{\prime\prime}\psi + 2\psi^{\prime}\sigma^{\prime}
+ 2\sigma^{\prime 2}\psi) - 4 V(\psi) \,=\, 0\,.
\label{8}
\eeq
Here the prime stands for the derivative with respect to the
conformal time, e.g.,
$$
\sigma^{\prime} \,=\,\frac{d\si}{d\eta} = a(t)\,
\frac{d\si}{dt}\,,
$$
while the derivative with respect to the physical time $t$ is
denoted as a dot. It terms of the physical time and adopting $\,k=0$,
we obtain the relation \beq -12 H_{0}^{2}\,\psi - 9 H_{0}\,
\dot{\psi} - 3 \,\ddot{\psi}\, \,=\, \,2 \,V\,. \label{9} \eeq
Using $\psi\,=\,-\kappa^{2} + \alpha\,R$ and taking into
account $R\,=\,-12\,H_{0}^{2}$ for the FRW metric, we get \beq
V\,=\,6\,H_{0}^{2}\,\kappa^{2} - 6\,H_{0}^{2}\,\alpha\,R \eeq and
finally \beq \Omega\,=\,6\,H_{0}^{2}\,\kappa^{2} \, .
\label{Om} \eeq On the other hand, starting from \beq S_{f}\, =\,
\int\,d^{4}x\,\sqrt{-g}\,f(R)
\,=\,\int\,d^{4}x\,\sqrt{-g}\,\Big[\Omega - \kappa^{2}\,R +
\frac{\alpha}{2}\,R^{2}\Big] \eeq we arrive at the equation \beq
\frac{1}{\sqrt{-g}}\,\frac{\delta\,S_{f}}{\delta\,\sigma} & = & 4
\Omega - 6 \kappa^{2}  e^{-2\sigma}(2 \sigma^{\prime \, 2} + 2
\sigma^{\prime\prime}) + 18 \alpha e^{-4 \sigma}(2
\sigma^{\prime\prime\prime\prime} - 12 \sigma^{\prime \, 2}
\sigma^{\prime\prime})=0\,. \eeq It is straightforward to check
that the solution $\sigma=H_0t=-\ln (H_0|\eta|)$ corresponds,
again, to the relation (\ref{Om}).


Finally, for the sake of completeness, let us address the
possibility of using conformal transformation to deal with
the metric-scalar theory.
It is well known fact that the theory (\ref{2}) can be easily
mapped into another one with the standard form of the scalar
kinetic term. We will give the corresponding treatment here just
for completeness and refer the reader to the review \cite{Faraoni}
for further details and (numerous, indeed) references.

Let us start from the conformal transformation
\beq
g_{\mu\nu}
\,\, \longrightarrow \,\,{\bar g}_{\mu\nu}\,=\,g_{\mu\nu}\cdot
e^{2\si(x)}
\label{1.7}
\eeq
in the action (\ref{2}). Simple
calculation yields the following result
\beq
S_c[g_{\mu\nu}e^{2\si},\,\psi] \,=\, \int d^4x \sqrt{-g}\;
\big\{\psi\,e^{2\si}\,\left[ R - 6(\na\si)^2 - 6\Box\si\right]
\,-\,e^{4\si}\, V(\psi) \big\}\,,
\label{1.3}
\eeq
where
$(\na\si)^2 = g^{\mu\nu}\pa_\mu\si\pa_\nu\si$. Let us choose
$\si$ such that $\,\psi\,e^{2\si}=-\ka^2$. Then the first term
$\psi\,e^{2\si}\,R$ coincides with the Einstein-Hilbert term, also
the third term $\, - 6\psi\,e^{2\si}\Box\si\,$ becomes a total
derivative which does not affect the equations of motion. In order
to provide the standard form of the kinetic term, we take
\beq
\ph=2\,\sqrt{3}\,\ka\,\si\,, \qquad\mbox{then}\qquad
\psi=-\ka^2\,\exp\big\{-\frac{\ph}{\sqrt{3}\,\ka}\big\}\,.
\label{1.4}
\eeq
The output looks like
\beq
S_{min}[g_{\mu\nu},\ph] = \int d^4x \sqrt{-g}\; \big\{ -\ka^2R
\,+\, \frac{1}{2}\,g^{\mu\nu}\pa_\mu\ph\pa_\nu\ph - U(\ph)\big\}\,,
\label{1.1}
\eeq
where the two potentials are related as
\beq
U(\ph)\,=\,e^{4\si}\,V(\psi)\,=\,\frac{\ka^4}{\psi^2}\,V(\psi)\,.
\label{1.6}
\eeq
The formulas (\ref{1.7}), (\ref{1.4}) and
(\ref{1.6}) are nothing but the change of variables in the action
(\ref{2}) which transform it into the action (\ref{1.1}).
Therefore we do not need to check the equivalence between two
actions by other means, e.g. examining the equations of motion.
Finally, let us note that the multiscalar case can be, in
principle, also treated by the conformal transformation, but this
transformation is not so easy as in the one-scalar case
\cite{Kaiser}.

\section{Modified gravity theory of a more general form}

We have presented a useful and simple prescription of mapping
theories (\ref{1}) into theories (\ref{2}) at the classical
level. This method can be generalized to the gravitational
actions which are more general than (\ref{1}). However, as we
will see in what follows,
in this case one needs more scalar fields. Some similar
results has been recently obtained in \cite{Felic} and
\cite{Felic-Tan}, but the derivation there looks rather different.

The method may be
especially useful for working out the dS-type solutions and,
therefore, is applicable for testing various models of modified
gravity, including the ones corresponding to quantum corrections.
Consider the following gravitational action
\beq
S_{gen} \,=\, \int d^{4}x\,\sqrt{-g}\,f(R,\,\Box R,\,C^2,
\,\widetilde{E})\,,
\label{s2-ini}
\eeq
where \ $\widetilde{E}\,=\,E - \frac{2}{3}\,\Box\,R$,
\ $\,E = R_{\mu\nu\al\be}^2 - 4 R_{\al\be}^2 + R^2$ \
is the Gauss-Bonnet topological term (Euler density) and
 $\,C^2=R_{\mu\nu\al\be}^2 - 2 R_{\al\be}^2 + (1/3)\,R^2\,$
is the square of the Weyl tensor. In view of the cosmological
applications, it proves more useful to consider \ $\widetilde{E}$
\ rather than  \ $E$.

We start by introducing a generalization of the action (\ref{1}),
\beq
S_{1}\,\,\,=\,\int\,d^{4}x\,\sqrt{-g}\,f(X_{i})
\qquad \mbox{where}\qquad
X_{i}\,=\,R,\,\Box R,\,\widetilde{E},\,C^{2}
\label{s1}
\eeq
in case of the action (\ref{s2-ini}), but the number of
invariants can be easily extended. For this end we
set \ $i\,\,\,=\,\,\,1,\,...\,,N$. Consider the dual action
\beq
S_{2} \,\,=\,\int\, d^{4}x\,\sqrt{-g}\,
\Big[X_{i}\psi^{i} - V(\psi^{i})\Big]\,,
\label{s2}
\eeq
where repeated indices imply summation, as usual.
Let us follow the same scheme which we applied in the previous
section. The equations for the $\psi^{i}$ have the form
\beq
X_{i} \,=\, \frac{\pa V}{\pa \psi^i}\,.
\label{29}
\eeq
We put them into (\ref{s2}), demanding equivalence
to the action (\ref{s1}),
\beq
S_{2} \,=\,
\int\,d^{4}x\,\sqrt{-g}\,\Bigl\{\psi^{i}
\frac{\pa V}{\pa \psi^i} - V(\psi^{i})\Bigl\}
\,=\, \int\,d^{4}x\,\sqrt{-g}\,f(X_{i})\,.
\eeq
Assuming
\beq
f(X_{i}) \,=\, X_{i}\psi^{i} - V(\psi^{i})\,
\label{equi}
\eeq
and taking partial derivatives with respect to $X^i$ in
(\ref{equi}), we arrive at
\beq
\frac{\partial f}{\partial X_k}
\,=\, \psi^k + X_i\,\frac{\partial \psi^i}{\partial X_i}
- \frac{\partial V}{\partial \psi^i}\,
\frac{\partial \psi^i}{\partial X_k}
\quad\Longrightarrow\quad
\psi^{k} \,=\,\frac{\partial f}{\partial X_k} \,,
\label{31}
\eeq
where we used Eq. (\ref{29}).
The formulas (\ref{29}) and (\ref{31}) show that we always have
\beq
X_i \,=\,
\frac{\partial\,V(\psi)}{\partial\,\psi_i}
\quad \mbox{and}\quad
\psi_k \,=\,
\frac{\partial\,f(X)}{\partial\,X_{k}}
\label{32}\,.
\eeq
After all, the prescription for deriving $V(\psi^i)$ is very
similar to the one described in the previous section and looks
as follows:

a) calculate \ $\psi^k = \frac{\partial\,f(X)}{\partial\,X_{k}}$;

b) solve these equations and find \ $X_k(\psi) \,=\,
\frac{\partial\,V}{\partial\psi^k}$;

c) integrate the last relations and find $V(\psi_k)$ up to the
additive constant;

d) fix this constant by the requirement that the actions
coincide in the corresponding limit (typically zero curvature).

One can note that this procedure can be applied also to the
non-Riemannian generalizations of GR, including the theory
of gravity with torsion.

Let us consider an example of how the equivalent metric-scalar
theory can be achieved. We start from the action (\ref{s2-ini})
with the function
\beq
f(R,\widetilde{E}) \,=\, F(R) \cdot \widetilde{E}\,,
\label{33}
\eeq
where $F(R)$ is an arbitrary function of scalar curvature,
which will be fixed later on. The equivalent action is
\beq
S_{2} \,=\,
\int d^4x\sqrt{-g}\,
\Bigl\{\psi R + \chi \widetilde{E} - V(\psi,\chi)\Bigl\}\,.
\label{34}
\eeq
Let us follow the prescription described above. The equations
\beq
\frac{\partial f}{\partial \widetilde{E}}
=\chi = F(R)
\,,\qquad
\frac{\partial f}{\partial R} = \psi
= \widetilde{E} F^{\prime}(R)
\eeq
can be solved with respect to the two scalar fields,
\beq
R = g(\chi)\,,
\qquad
\widetilde{E} = \frac{\psi}{F^{\prime}(R)}
\quad \Longrightarrow \quad
\widetilde{E} = \frac{\psi}{F_{g}^{\prime}(g(\chi))}\,.
\eeq
On the other hand, we have inverse functions
\beq
R = \frac{\partial V}{\partial \psi} = g(\chi)
\,,\qquad
\widetilde{E}
\,=\, \frac{\partial V}{\partial \chi}
\,=\, \frac{\psi}{F_{g}^{\prime}(g(\chi))}\,.
\eeq
Then
\beq
V(\psi,\chi)
\,=\, \int\,g(\chi)\,d\psi + g_{1}(\chi)
\,=\, g_{1}(\chi) + \psi \,g(\chi)
\label{35}
\eeq
and
\beq
V(\psi,\,\chi)
\,=\, \int d\chi \,\frac{\psi}{F_{g}^{\prime}(g(\chi))} + g_{2}(\psi)
\,=\, \psi\,\int \frac{d\chi}{F_{g}^{\prime}(g(\chi))} + g_{2}(\psi)\,.
\label{36}
\eeq
If we compare the two forms of the potential function (\ref{35})
and (\ref{36}), it becomes clear that \ $g(\chi)$ \ satisfies the
functional equation
\beq
g(\chi)
\,=\, \int \frac{d\chi}{F_{g}^{\prime}(g(\chi))} \,+\,C\,.
\label{funk}
\eeq
and, moreover,
\beq
g_2(\psi) - C\,\psi \,=\, g_1(\chi) \,=\,C_1\,=\,const\,.
\eeq

It does not look possible to advance further, so let us
take a more concrete form of \ $F(R)$. Consider first a very
simple case
$\,f (R,\,\widetilde {E}) = - \Omega + R \, \widetilde{E}$.
Making the same steps as in the general case, we obtain
\beq
F(R) \,=\,R\,\Longrightarrow\,\chi\,=\,R
\,,\qquad g(\chi) \,=\, \chi\,,
\qquad F^{\prime}(R) \,=\, 1\,.
\nonumber
\eeq
Using (\ref{35}) and (\ref{36}), one obtain
\beq
V(\psi,\,\chi) \,=\,
g_{1}(\chi) + \psi\chi \,=\,
\psi\int\,d\chi + g_{2}(\psi) \,=\, \psi\chi + g_{2}(\psi)\,.
\nonumber
\eeq
It is easy to see that in this case
\ $g_{1}(\chi) = g_{2}(\psi) = C$, so we get
\beq
V(\psi,\chi) = \psi \chi + C\,.
\label{REpot}
\eeq
Finally, inserting the relations $\chi=R$ and $\psi=\widetilde{E}$
into (\ref{REpot}), one can easily verify that $C\,=\Omega$. This
result can be also checked by inspecting exponential solutions
in the two cases. We avoid to bother the reader with the details
of this verification, but just note that its output is positive.

Consider a bit more complicated example when
\beq
f(R,\widetilde{E})
\,=\,\Omega - \kappa^{2}\,R + \be\,\widetilde{E}\,\ln\,
\Big(1 + \frac{R}{R_{0}}\Big)\,,
\label{cov-E}
\eeq
where $\be$ is some constant and $R_0$ is a reference value
for the scalar curvature.
The expression (\ref{cov-E}) can be seen as a part of the
renormalization
group corrected vacuum action, where the renormalization
group parameter $\mu^2$ is associated to the scalar
curvature (see, e.g., \cite{Poimpo} for further details
and references).

Following the footsteps of the previous examples, we derive
\beq
\frac{\pa\,f(R,\widetilde{E})}{\pa\,\widetilde{E}}
\,=\,\chi\,=\,\be \ln \Big(1 + \frac{R}{R_{0}}\Big)
\quad \Longrightarrow \quad
R\,=\,R_{0}\Big(e^{\chi/\be} - 1 \Big)
\eeq
and
\beq
\frac{\pa\,f(R,\widetilde{E})}{\pa\,R}
\,=\,\psi\,=\,\frac{\be\,\widetilde{E}}{R + R_{0}} - \ka^{2}
\quad \Longrightarrow \quad
\widetilde{E}\,=\,\frac{R_0}{\be}\,(\psi + \ka^{2})
\,e^{\chi/\be}\,.
\eeq
At the next stage we find
\beq
\frac{\pa\,V(\psi,\chi)}{\pa\,\psi}
\,=\,
R\quad\Longrightarrow\quad
V(\psi,\chi) &=& g_{1}(\chi)
+ R_{0}\,\psi\,e^{\chi/\be} - R_{0}\,\psi\,.
\nonumber
\\
\frac{\pa\,V(\psi,\chi)}{\pa\,\chi}
\,=\,\widetilde{E}
\quad\Longrightarrow\quad
V(\psi,\chi) &=& g_{2}(\psi) + R_{0}\,\psi\,e^{\chi/\be}
+ \ka^{2}\,R_{0}\,e^{\chi/\be}\,.
\nonumber
\eeq
Using these two expressions it is easy to figure out that
$$
g_{1}(\chi)\,=\,\ka^{2}\,R_{0}\,e^{\chi/\be}+C
\quad
\mbox{and}
\quad
g_{2}(\psi)\,=\,- R_{0}\,\psi+C\,.
$$
Finally, we arrive at the potential
\beq
V(\psi,\chi)\,=\,R_{0}\,\psi\,\Big(
e^{\chi/\be} - 1\Big)
+ \ka^{2}\,R_{0}\,e^{\chi/\be} - \Om - \ka^2 R_0\,,
\label{Pot-be}
\eeq
where the constant \ $C =-\Om - \ka^2 R_0$ has been fixed 
following the same method which we used in the previous cases.

One can consider more complicated expression for the
covariant Lagrangian,
\beq
f(R,\widetilde{E},C^2)
\,=\,\Omega - \kappa^{2}\,R + \be\,\widetilde{E}\,\ln\,
\Big(1 + \frac{R}{R_{0}}\Big)  + \be_1\,C^2\,\ln\,
\Big(1 + \frac{R}{R_{0}}\Big)\,.
\label{cov-EB}
\eeq
At this point one can make an important observation. From the
first sight, the equivalent Lagrangian for this case should
have three auxiliary fields, because there are three structures
$R$, $\widetilde{E}$ and $C^2$. At the same time, the problem
of reducing the theory (\ref{cov-EB}) is essentially
equivalent to the one of the theory (\ref{cov-E}),
with the $\be\,\widetilde{E}$ traded by the combination
$\be\,\widetilde{E} + \be_1\,C^2$. Obviously, in this
case we need only two auxiliary fields and not three of
them. In other words, in this case the number of necessary
auxiliary fields is smaller than the one which could be
thought by just counting the number of the structures
$X_i$ in the starting action. This example shows that it
would be
interesting to have a general criteria for establishing
an exact number of necessary auxiliary fields for a given
initial function $f(X_i)$.

The problem of our interest is very close to the one which
is typical for the transition from Lagrange to Hamiltonian
formalism in the theory with constraints
\footnote{Recently, this similarity has been noted in
\cite{SalHin}.} (see, e.g., well-known books
\cite{gt,ht} for introduction purposes). Indeed, it is
analogous to the passage from the Lagrangian description,
with no explicit dependence on the coordinates, to the
Hamiltonian one. In this case $f(X^i)$ and
\ $X^ i\psi_i-V(\psi_i)$ \
play the roles of the Lagrangian and the Hamiltonian
respectively, where the ``momenta'' are defined by
\beq
\psi_i \equiv \frac{\partial f}{\partial X^i}\,.
\label{mom}
\eeq
Finally, the quantities of $\{ X^1,X^2,X^3\}$ play the
roles of ``velocities". For the specific case of eq.
(\ref{cov-EB}), these equations have the form
\beq
\label{psi1}
\psi_1 &=& - \kappa^2 +  \frac{\beta \tilde E
+ \beta_1 C^2} {R_0 + R},
\\
\psi_2 &=& \beta \ln \left( 1 + \frac R {R_0}\right),
\\
\psi_3 &=& \beta_1 \ln \left( 1 + \frac R {R_0}\right),
\eeq
from what we directly infer the presence of a constraint,
which is a dependence relation between the $\psi_i$  given by
\beq
\phi \equiv \beta_1 \psi_2 - \beta \psi_3 \equiv
\beta_1 \chi - \beta \chi_1 = 0\,.
\label{constphi}
\eeq
Since the constraint $\phi(\psi_i)$ comes directly from the
definition of the ``momenta", it is classified as a primary
constraint \cite{gt, ht}. We note that the equation
$\phi(\psi_i)=0$ defines a surface in the space
$\{\psi_1, \psi_2, \psi_3\}$. It should be stressed that
this constraint only represents a restriction in the
``momenta" space, while in the ``velocity" space
$\{ X^1,X^2,X^3\}$ the constraint does not lead to any
restrictions, since if the $\psi_i$ are written
as functions of the $X^k$, the constraint $\phi(\psi_i)$
becomes the function $\phi(\psi_i(X^k))$, which is
identically null.

If this were a typical Hamiltonian problem, one would evaluate
the evolution of the primary constraints (e.g., the single one
in the example considered above)
in search for further constraints. However our present problem
is simpler, because it does not involve any dynamics. Hence
only the constraints with no relation to dynamics are relevant
here, which are the primary ones.

\vspace{.1in}
Before proceeding towards the determination of the potential
$V$, we remark here on the relation between the Hessian matrix
\beq
\left( \frac{\partial^2 f}{ \partial X^i \partial X^j} \right)
\nonumber
\eeq
and the presence of constraints. Firstly, if the Hessian is
non-singular, the inverse function theorem guarantees that
(at least locally) one can use the definition (\ref{mom}) to
express the $X^k$ as a function of the $\psi_i$, and thus
no constraint is expected. However the Hessian can turn out
to be degenerate. For instance, in the case of
(\ref{cov-EB}), this $3\times 3$ matrix is a singular matrix
of rank two (for $R \not = - R_0)$, namely
\beq
\left( \frac{\pa^2 f}{\pa X^i \pa X^j} \right)
= \frac{1}{(R + R_0)^2}\begin{pmatrix}
- \tilde E \beta + C^2 \beta_1  &  \beta  & \beta_1
\\  \beta    &  0  &  0
\\  \beta_1  &  0  &  0
\end{pmatrix}\,.
\label{Hessian}
\eeq
Therefore, it has a single independent zero-mode (i.e., an
eigenvector whose corresponding eigenvalue is zero). Indeed,
\beq
\nu = \begin{pmatrix} 0 &  \beta_1 & - \beta \end{pmatrix}\,
\eeq
can be promptly identified as the single linearly independent
zero-mode of the Hessian (\ref{Hessian}). Let us note that we
choose to work with the zero-modes that multiply the Hessian
matrix by the left.

To conclude this introductory part, we note that each independent
zero-mode generates an independent constraint. In particular, by
multiplying the zero-mode $\nu$ on both sides of the definition
(\ref{mom}), one finds the same constraint (\ref{constphi}).
Afterwards we will show that each independent constraint leads
to an independent zero-mode of the Hessian matrix. We note that
this simple relation between zero-modes and constraints does not
have a counterpart in general Hamiltonian problems with
constraints, in particular since the corresponding
zero-modes may depend on "coordinates" there (in the present
problem, we are considering the analogous Hamiltonian problem
in which the Lagrangian only depend on the ``velocities" $X^i$).
\vskip 2mm

In the presence of constraints, one cannot use the first
relation of (\ref{32}) to find $V$, since this relation is
not valid in the presence of constraints. Namely, consider
the variation of $V$  in the constraint surface (i.e., in
the surface $\phi = 0$),
\beq
\delta V = \delta( \psi_i X^i  - f) = X^i \delta \psi_i
+ \left( \psi_i  - \frac{\partial f}{\partial X^i} \right)
\delta X^i =  X^i \delta \psi_i \,.
\eeq
In the last step above, we used the definition of the momenta
(\ref{mom}). The previous equation shows that $V$ can be
written as a function of $\psi$ alone, even if constraints
are present. Thus, using the last equality,
\beq
\left( \frac{\partial V}{\partial \psi_i} - X^i\right)
\delta \psi_i = 0\,.
\eeq
Now, using the Theorem 1.2 of \cite{ht}, we find the extension
of the first relation of (\ref{32}) to the constrained case,
\beq
X^i = \frac{\partial V}{\partial \psi_i}
+ \lambda_m \frac{\partial \phi_m}{\partial \psi_i}\,.
\label{Rconst}
\eeq
In this formula $\lambda_m$ are Lagrange multipliers and
$\phi_m$, with $m=1,2,...,M$, are all the constraints of the
model under consideration. In the particular case of
(\ref{cov-EB}) we have a single constraint, i.e. $M=1$. The
introduction of these multipliers is necessary since the
relation between the $X_i$ and the $\psi^k$ has to be
extended in order to become invertible.

Now we are in a position to discuss the method of
constructing potential $V(\psi_i)$ in the case of a theory
with constraints. The integration method which was employed
previously can be extended to this case. For instance, in
the example of $f(R,\tilde E, C^2)$ theory (\ref{cov-EB})
one can solve the  definition of $\psi_i$ and arrive at
\beq
\label{Rchoice}
R &=& R_0 \left(  e^{\chi/\beta} - 1\right ),
\\
\tilde E & =& \frac{R_0}{\beta} e^{\chi/\beta}
(\psi + \kappa^2) - \frac{\beta_1}{\beta} C^2.
\eeq
Hence, from (\ref{Rconst}) and  (\ref{constphi}) we find
\beq
\label{Rfirst}
&& R_0 \left(  e^{\chi/\beta} - 1\right )
=  \frac{\partial V}{\partial \psi},
\\
&& \frac{R_0}{\beta} e^{\chi/\beta} (\psi + \kappa^2)
- \frac{\beta_1}{\beta} C^2
=  \frac{\partial V}{\partial \chi} + \lambda \beta_1,
\\
\label{c2third}
&&
C^2 = \frac{\partial V}{\partial \chi_1} - \lambda \beta
\eeq
The first equation can be straightforwardly integrated,
but the second cannot, since we do not know how to express
both $C^2$ and $\lambda$ as functions of the $\psi_i$.
Nevertheless, the Lagrange multiplier is still free, thus
we can set it in such a way that the $C^2$ term disappear,
namely $\lambda = - C^2/\beta$. Consequently, the third
equation (\ref{c2third}) becomes
$$
\frac{\partial V}{\partial \chi_1} = 0\,.
$$
After that the expression for $V$ can be integrated,
and $V(\psi_i)$ can be found using procedures similar
to that used in the unconstrained case. One should note
that the number of auxiliary fields in this procedure
is smaller than the number $N$ of the $X_i$ structures
in the initial $f(X_i)$ theory. For example, in the
(\ref{cov-EB}) case, albeit we started from the $f(X_i)$
which depends on three independent quantities, the
corresponding potential $V$ only depends on the two
independent scalar fields.

The form of the potential $V$ can have some impact
on the physical consequences of a given theory $f(X_i)$.
Hence, before concluding this example, we present
$V$ in a more general form. Let us start, as usual, from
the simple example. In Eq. (\ref{Rchoice}) we have
made a choice of selecting the particular expression of
$R$ as a function of $\chi$ and $\chi_1$. At the same
time one can express $R$ in a more general form,
\beq
R =  \frac{R_0}{1 + \xi}\, \left[  e^{\chi/\beta} - 1
+ \xi  \big(  e^{\chi_1/\beta_1} - 1\big)\right]\,,
\label{R-xi}
\eeq
where $\xi$ is an arbitrary real number different from
$-1$. The choice in (\ref{Rchoice}) corresponds to
$\xi = 0$. From the more general version (\ref{R-xi})
and Eq. (\ref{psi1}) one can easily obtain
\beq
\beta \tilde E + \beta_1 C^2 = R_0
\frac{(\psi + \kappa^2)}{\xi + 1} \left( e^{\chi/\beta}
+ \xi e^{\chi_1/\beta_1} \right)\,.
\eeq
Now, instead of attempting a direct integration of $V$
(as in the non-constrained case), we express it on the
constraint surface $\phi=0$, as
\beq
V = \psi_i X^i - f &=& \psi_i X^i
- \left[ \Omega - \kappa^2 R_0  \frac{(e^{\chi/\beta} -1)
+ \xi \, (e^{\chi_1/\beta_1}-1)}{\xi + 1}
+ \tilde E \chi + C^2 \chi_1\right]
\\
&=&  R_0  (\psi + \kappa^2) \frac{e^{\chi/\beta} -1 + \xi\, 
(e^{\chi_1/\beta_1}-1)}{\xi + 1} - \Omega
\\
\label{Vfff}
&=& R_0  (\psi + \kappa^2) \left( e^{\chi/\beta}-1 \right)
- \Omega.
\eeq
As one should expect, on the constraint surface $V$ is
independent on the value of $\xi$. Nevertheless, a choice of
$\xi=1$ or $\xi=0$, for instance, might have computational
advantages one over the other. Continuing the
integration of $V$ as in the previous method, it is 
straightforward to check that the final answer is given by 
(\ref{Vfff}).
\vskip 1mm

With the experience which we just gained from the example
considered above, it is not hard to guess that, in general,
there is a straight relation between the number of
independent zero-modes of the Hessian and the number of
constraints. Consider the case in which there are $M$
independent constraints given by
\beq
\phi_m(\psi_i) = 0\,,
\label{Phi}
\eeq
with $m = 1,2,...,M$. We assume that the constraints are
expressed such that the gradients of the constraints are
linearly independent on the constraint surface (for more
details, see the regularity conditions on how to express
the constraints \cite{ht}).

Since the primary constraints $\phi_m(\psi)$ are identically
null when expressed as functions of the $X^i$ variables,
\beq
\frac{\pa \phi_m(\psi_k(X^j))}{\pa X^i}
=  \frac{\pa \phi_m (\psi_k)}{\pa \psi_j}\,
\frac{\pa^2 f}{\pa X^i \pa X^j}\, = 0 \,.
\label{Phi-1}
\eeq
Then, for each independent constraint $\phi_m(\psi_i)$
there is a corresponding independent zero-mode of the
Hessian given by
\beq
(\nu^j)_m = \frac{\partial \phi_m }{\partial \psi_j}.
\eeq
In other words, upon transforming a Lagrangian given by
$f(X^i)$, with $i=1,2,...,N$, into an equivalent one given
by \ $\psi_i X^i - V(\psi_i)$,
the number of independent auxiliary scalar fields that
appear in the potential $V(\psi_i)$ is equal to the rank
of the Hessian of $f(X^i)$.
\vskip 2mm

In particular, if $f$ depends on $R$ and $E$ only, the condition
of the degeneracy of the Hessian matrix reduces to \beq
f_{RR}f_{EE}-f_{RE}^2=0 \, . \eeq The principal difference between
the behaviour of solutions in this special case and in the general
one has been already noticed when studying small inhomogeneous
perturbations on a Friedmann-Robertson-Walker (FRW) background
\cite{Felic}.

A relevant general observation is that we only need the
potential $V$ computed on the constraint surface, since
in general
\cite{ht}
\beq
V = V|_{\phi=0} + \lambda \phi,
\eeq
where $V|_{\phi=0}$ is the potential on the constraint surface,
but the (``primary") constraint $\phi$ identically vanishes when
expressed as a function of $R's$. Thus, it does not make
any difference whether one uses either $f = X^i \psi_i - V$ or
$f = X^i \psi_i - V|_{\phi=0}$.

\section{Example: conditions for exponential inflation}

As an illustration of the equivalence theorem from the previous
section, let us formulate the conditions for the existence of a
strictly exponential cosmological solution. Consider the action
\beq 
S_{eq} = \int d^{4}x\,\,\sqrt{-g}\,\,
\{\psi_{1}R + \psi_{2}\Box R + \psi_{3}C^{2} 
+ \psi_{4}(E-\frac{2}{3}\Box R) 
- V(\psi_{1},\psi_{2},\psi_{3},\psi_{4})\} 
\label{E1} 
\eeq 
and the metric 
\beq 
ds^{2} = g_{\mu\nu}dx^{\mu}dx^{\nu} =
a^{2}(\eta)(d\eta^{2} - dl^{2})\,, \qquad a(\eta)=e^{\si(\eta)}\,,
\label{E2} 
\eeq 
where 
\beq
dl^{2} = \frac{dr^{2}}{1 - kr^{2}} + r^2d\th^2 + r^{2}\sin^{2}\th
d\phi^{2}\,. 
\label{E3} 
\eeq 
The conformally transformed metric has the form 
\beq 
\bar{g}_{\mu\nu} \,=\, \diag \Big(1,\,
-\frac{1}{1 - kr^2},\, -r^2,\, -r^2\sin^2\th \Big) 
\label{E4} 
\eeq
In this section we will restrict our attention to the spatially
flat $k=0$ case. Then an exponential solution produces the de
Sitter space-time (it is not so for $k\not =0$).

Let us consider the variational derivatives
$\,{\delta S_{eq}}/{\delta \psi_{i}}$.
\beq
R
&-& \frac{\pa}{\pa \psi_{1}}V(\psi_{1},\psi_{2},\psi_{3},\psi_{4}) 
= 0
\nonumber
\\
\Box R &-&
\frac{\pa}{\pa \psi_{2}}V(\psi_{1},\psi_{2},\psi_{3},\psi_{4})= 0\,
\nonumber
\\
C^{2}  &-&  \frac{\pa}{\pa
\psi_{3}}V(\psi_{1},\psi_{2},\psi_{3},\psi_{4})= 0\,,
\nonumber
\\
(E-\frac{2}{3}\Box R)
&-&
\frac{\pa}{\pa \psi_{4}}V(\psi_{1},\psi_{2},\psi_{3},\psi_{4}) = 0\,.
\label{E7}
\eeq
Furthermore, we need the equation for the
metric, which can be obtained by taking the derivative of the action
with respect to $\sigma$, $\,\de S_{eq}/\de \si=0$. This gives
\beq
&-&  3e^{-4\si}\,\big(4\si^{\prime 3}
\psi_{2}^{\prime}
- 2\psi_{2}^{\prime\prime}\sigma^{\prime 2}
- 4\psi_{2}^{\prime}\sigma^{\prime}\sigma^{\prime\prime}
+ \psi_{2}^{\prime\prime\prime\prime}
+ 4\psi_{2}^{\prime\prime}\sigma^{\prime\prime}
+ 2\psi_{2}^{\prime}\sigma^{\prime\prime\prime}\big)
\nonumber
\\
&-& 3e^{-2\sigma}(\psi_{1}^{\prime\prime} +
2\sigma^{\prime\prime}\psi_{1} + 2\psi_{1}^{\prime}\sigma^{\prime}
+ 2\sigma^{\prime 2}\psi_{1}) +
2e^{-4\sigma}\psi_{4}^{\prime\prime\prime\prime} \,-\,
2V(\psi_{1},\psi_{2},\psi_{3},\psi_{4}) = 0\,.
\label{E9}
\eeq
If we assume that there exists an exponential (in terms of physical 
time) solution $a=a_{0}e^{H_{0}t}$, it can be inserted into the 
equations (\ref{E9}) with the following output:
\beq
\frac{\pa V}{\pa \psi_{1}} = -12 H_0^2\,,\qquad
\frac{\pa V}{\partial \psi_{2}} = 0\,,\qquad
\frac{\pa V}{\pa \psi_{3}} = 0
\quad\mbox{and}\quad
\frac{\pa V}{\partial \psi_{4}} = 24H_{0}^{4} \,.
\label{dS1}
\eeq
Furthermore, the equation for the conformal factor has the form
\beq
&-&  4V - 24 H_{0}^{2}\psi_{1} - 18 H_{0}\dot{\psi_{1}}
- 6 \ddot{\psi_{1}} - 72\dot{\psi_{2}}H_{0}^{3}
- 78 H_{0}^{2}\ddot{\psi_{2}}
\nonumber
\\
&+&  36 H_{0}\dot{\ddot{\psi_{2}}} +6 \ddot{\ddot{\psi_{2}}}
+ 24 H_{0}^{3}\dot{\psi_{4}} + 44 H_{0}^{2}\ddot{\psi_{4}}
+ 24 H_{0} \dot{\ddot{\psi_{4}}} =0
\label{dS5}
\eeq
Finding of dS-like solutions in this way implies resolving 
the system of equations (\ref{dS1}), (\ref{dS5}), but it is
not clear whether this can be done in a general form. It
should be noted that the {\it r.h.s.} of the equations
(\ref{dS1}) - (\ref{dS5}) are valid only assuming that we
have already used some (unknown) solutions for the
auxiliary scalars $\psi_k$ there.

Let us try another approach for de Sitter solutions. 
Consider first, as a heat-up exercise, the theories with 
the actions (\ref{1}) and (\ref{2}). We know from the 
Sect. 2 that (\ref{1}) and (\ref{2}) are equivalent, 
provided that the functions $\psi = f^\prime_R(R)$ and 
$R = V_\psi^\prime(\psi)$ are inverse functions. Let us 
use this fact to obtain the criterion of dS-like solution 
for the theories (\ref{1}) and (\ref{2}). The dS solution 
means, in the new frame, that 
\beq
R_{\mu\nu}=\frac14\,Rg_{\mu\nu}
\quad
\mbox{and}
\quad 
R_{\mu\nu\al\be}=\frac{1}{12}
\,R\big(g_{\mu\al}g_{\nu\be}-g_{\nu\al}g_{\mu\be}\big)
\,,\quad \mbox{also}\quad R=const\,.
\label{ad0}
\eeq 

How can we see whether the solution of the form (\ref{ad0}) 
is possible or not for the given theory? In case of (\ref{1}) 
one can take variation with respect to the metric, 
$\de g_{\mu\nu}=h_{\mu\nu}$, and arrive at the equation 
\beq
\frac12\,f\,g_{\mu\nu} - R_{\mu\nu}f^\prime_R + 
(\na_\mu\na_\nu - g_{\mu\nu}\Box)f^\prime_R = 0 \,.
\label{ad1}
\eeq
Inserting (\ref{ad0}) into the last equation, we arrive at 
the well known algebraic equation which roots $R=const$ give 
us dS solutions of $f(R)$ gravity: 
\beq
\frac12\,g_{\mu\nu}\,f - \frac14\,Rg_{\mu\nu}f^\prime_R = 0 
\quad \Longrightarrow\quad
Rf^\prime_R = 2f\,.
\label{ad2}
\eeq

Now we can do the same for the (\ref{2}) version of the 
same theory. The equations equivalent to (\ref{ad1}) in 
this case have the form 
\beq
\frac12\,g_{\mu\nu}\,
\big(\psi R - V\big) - \psi R_{\mu\nu} = 0 
\,,\quad \mbox{also}\quad R=V^\prime_\psi\,.
\label{ad3}
\eeq
Inserting (\ref{ad0}) into (\ref{ad3}), we arrive at the 
algebraic equation for constant values of  $\psi$ 
at all possible dS solutions:
\beq
\psi\, V^\prime_\psi = 2V\,.
\label{ad4}
\eeq
It is fairly easy to see that Eq. (\ref{ad4}) is nothing else
but the mapping of the final relation in Eq. (\ref{ad2}). For 
this end one has to just use our main relations 
$\psi = f^\prime_R(R)$ and $R = V_\psi^\prime(\psi)$ in 
(\ref{ad2}).

One can note that the relation (\ref{ad4}) plays exactly the 
same role for the theory (\ref{2}) as the relation (\ref{ad2}) 
does play for the theory (\ref{1}). 

The next task is to obtain similar relations for the theories
(\ref{s1}) and (\ref{s2}).
One can immediately notice that under the conditions (\ref{ad3})
we have $X_1=\Box R =0$ and $X_3=C^2=0$. Therefore the existence 
of the solution of the form (\ref{ad3}) concerns only the 
dependence of $f(X_i)$ on $X_2$ and $X_4$ in one case and the 
dependence of $V(\psi_i)$ on $\psi_2$ and $\psi_4$ in another 
case. In all cases we can also consider $E$ instead of $\tilde{E}$.

First we deal with Eq. (\ref{s1}). In taking variations 
of the metric we have to remember that after that we shall 
integrate by parts and then use the conditions (\ref{ad3}).
Therefore, all covariant derivatives, either acting on the 
variation of the metric $h_{\mu\nu}$ or on curvature tensor
components, can be safely neglected. In this way we obtain 
\beq
\de S_{1}\,=\,\int\,d^{4}x\,\sqrt{-g}\,\Big\{
\frac12\,hf + f^\prime_R\cdot \de_hR + f^\prime_E\cdot \de_hE
\Big\}\,.
\label{ad5}
\eeq
A very simple calculations give 
\beq
\de_hR\Big|_{dS}=-\frac14\,Rh
\,,\quad \mbox{and}\quad
\de_hE\Big|_{dS}=-\frac{1}{12}\,R^2h=-\frac{1}{2}\,Eh\,,
\eeq 
where we denoted $h=h^\mu_\mu$. Then, for all dS solutions
of this theory, the constant invariants $R$ and $E$, related
by the consistency condition $E=R^2/4$ in this case, should also
satisfy the "on-shell" algebraic equation
\beq
f \,=\, \frac12\,Rf^\prime_R \,+\, Ef^\prime_E\,.
\label{ad6} 
\eeq 
The last equation is a direct generalization of (\ref{ad2}) 
and has the same theoretical status for the more general theory
(\ref{s1}) as (\ref{ad2}) has for the theory (\ref{1}). It
follows from expressions presented in e.g. \cite{Carroll05,Felic}
(though we have been unable to find a paper where it was written 
explicitly).

The next step is to obtain the extension of Eq. (\ref{ad4})  
for the more general case of the theory (\ref{s2}). We take 
all three necessary variations and get
\beq
\frac{\de S_2}{\de \psi_1} &=& 0\quad \Longrightarrow\quad
R=V^\prime_1=\frac{\pa V}{\pa \psi_1}\,,
\nonumber
\\
\frac{\de S_2}{\de \psi_4} &=& 0\quad \Longrightarrow\quad
E=V^\prime_4=\frac{\pa V}{\pa \psi_4}\,,
\nonumber
\\
\left.\frac{\de S_2}{\de h_{\mu\nu}}\right|_{dS} 
&=& 0\quad \Longrightarrow\quad
\frac12\,\big(V-\psi_2R-\psi_4E\big)=
-\frac14\,\psi_2R -\frac12\,\psi_4E\,,
\label{ad7}
\eeq
It is easy to see from Eq. (\ref{ad7}) that one equation for
constant values of $\psi_i, \ i=1,2,3,4$, at the de Sitter solutions  
(\ref{ad0}) has the same form (\ref{ad4}) (with $\psi_1$ instead of 
$\psi$ and $V$ depending on all $\psi_i$) even if the Gauss-Bonnet 
and other terms are present (of course, this does not 
mean that these terms play no role here). 

The other three algebraic "on-shell" equations follow from (\ref{dS1}) 
by excluding $H_0$ (still to be found) and using the
relation between $R$ and $E$ for a dS solution: 
\beq
\frac{\pa V}{\pa \psi_{1}} = 0\,,\qquad
\frac{\pa V}{\pa \psi_{3}} = 0 \,, \qquad
\frac{\pa V}{\partial \psi_{4}} 
=  \frac{1}{4}\left(\frac{\pa V}{\partial \psi_{2}}\right)^2\,.
\label{dS2}
\eeq

\section{String-inspired case}

Now let us try to generalize the consideration given above to
more general gravitational actions containing more
complicated terms constructed from the scalar curvature $R$, the 
Ricci tensor $R_{\mu\nu}$ and the Riemann tensor 
$R_{\mu\nu\alpha\beta}$. In general, the
corresponding theories can not be reduced to the metric-scalar
models. For example, the $R^2_{\mu\nu\alpha\beta}$-type actions
involve higher derivatives not only in the spin-zero sector but
also in the spin-two one. Hence one can expect that the
reduction to second order equations would require introduction of 
tensor compensating fields. This is definitely true in general, 
however there is an interesting possibility yet. Let us consider 
a special space-time metric with a restricted number of degrees 
of freedom, such as the cosmological FRW one. In this case we 
have only one component of the metric -- the scale (conformal) 
factor, which depends on a single variable (e.g. conformal time). 
Then we meet a much simpler situation 
than in the general case, because the tensor structure of the
$f(R_{\mu\nu\alpha\beta})$ action becomes irrelevant. It might
happen that the reduction to the metric-scalar theory will be
possible in this case. Indeed, this reduction concerns only the
dynamics of the conformal factor of the metric. After this
dynamics is described in terms of an appropriate metric-scalar
theory, one has to explore other, more complicated aspects (e.g.
metric perturbations) in the framework of the original
higher-derivative theory. However, the equivalence with the
metric-scalar model may be a useful tool for dealing with a
homogeneous and isotropic cosmological solution. Therefore, it
deserves our attention.

Let us consider an effective low energy action of (super)string
theory (see, e.g. \cite{GSW}) depending only on metric. For the
sake of simplicity we assume that the dilaton and effective
fermionic degrees of freedom (after compactification to $d=4$,
because we are interested in the four-dimensional physics here)
are all in the vacuum state and only metric possesses dynamical
behaviour. This assumption is perfectly consistent with the known
form of the gravitational interaction which corresponds to General
Relativity (GR), the theory where only massless graviton is
propagating.

In the lowest, first order in the string parameter
$\alpha^\prime$, we meet the standard Einstein-Hilbert action for
gravity. In the next order in $\alpha^\prime$ we meet a set of
higher derivative terms, namely \beq R_{\mu\nu\al\be}^2\,,\qquad
R_{\mu\nu}^2 \qquad \mbox{and} \qquad R^2. \label{high} \eeq It
has been noticed 25 years ago by Zwiebach \cite{zwei} (see also
consequent investigation of the problem in \cite{dere}) that the
choice of the background string metric can be always done in such
a way that the higher order corrections do not generate unphysical
propagating massive ghosts - a typical phenomenon for a wide class
of higher derivative gravity theories \cite{stelle,highderi}. In
the second order in $\alpha^\prime$, the ghosts do not show up if
the higher derivative terms (\ref{high}) enter the following
combinations 
\beq 
E = R_{\mu\nu\al\be}^2 - 4R_{\mu\nu}^2 + R^2
\,,\qquad R^2\,. 
\label{high-end} 
\eeq
Possible metric reparametrizations in the ${\cal
{O}}\big({\al^\prime}^2\big)$ order have the form
\beq 
g_{\mu\nu} \longrightarrow g'_{\mu\nu} = g_{\mu\nu} + \al
'\left(x_1\,R_{\mu\nu} + x_2\,R\, g_{\mu\nu}\right) + ...\,,
\label{repar} 
\eeq 
where $x_{1,2}$ are  arbitrary parameters. The
same procedure can be used in higher orders. The above
transformations can change coefficients (in particular eliminate
completely) of all those terms which depend on the Ricci tensor or
on the scalar curvature $R$, only those terms which are
constructed exclusively from the Riemann tensor may not be
modified. The next question is what are the physical constraints
for those terms which can be modified. As we have already seen
above, the $R^2$ term can be traded for a scalar field. The first
term in the last expression is nothing else but the integrand of
the Gauss-Bonnet topological invariant (Euler characteristics) of
the space-time manifold. It is remarkable that this term does not
influence the dynamics of the universe. Hence one can completely
eliminate the relevant ${\cal O}(R^2)$ corrections to the string
effective action by means of the metric reparametrization. At the
same time, this procedure is not uniquely defined. If we require
the absence of unphysical ghosts, we may eliminate or not the
relevant $R^2$ term depending on our own will. The related
ambiguity may affect the cosmological solutions \cite{maroto} and
can not be fixed without experimental verification (see also
corresponding discussion for the case of the gravity with dilaton
and torsion in \cite{dere} and \cite{torsi}). The $R+R^2$-type
action corresponds to the choice of the metric parametrization
described above.

In the next order in $\alpha^\prime$, we meet corrections which
are cubic in curvature tensor and also the $R\Box R$-type terms.
Since the latter term generically leads to the appearance of a
ghost scalar \cite{GSS,ovrut}, we investigate the most general
metric parametrization in this given order of the former term
only. The corresponding effective action has the form 
\beq 
S_3 = \int d^4 \sqrt{-g} \,\left\{\, x_1R^3 + x_2RR_{\mu\nu}
R^{\mu\nu} + x_3RR_{\mu\nu\al\be}R^{\mu\nu\al\be} \right. 
\nonumber
\\
\left.
+ x_4 R_{\mu\nu}R^{\nu\al}R_\al^\mu
+ x_5 R_{\mu\nu\al\be}R^{\al\be\rho\la}R^{\mu\nu}\,_{\rho\la}\,
\right\}.
\label{3}
\eeq

Let us now consider the possibility of reduction of the
above theory (\ref{3}) to the
metric-scalar model in case of the special conformally flat
metric $g_{\mu\nu}={\bar g}_{\mu\nu}\,a^2(\eta)$. Here
$\eta$ is the conformal time and ${\bar g}_{\mu\nu}$ is
the time-independent homogeneous and isotropic metric (\ref{E4}).
For our purposes it is better to use the variable $\si(\eta)$,
defined in (\ref{E2}). The transformation rules
for the curvature scalar and tensors has the form
\beq
R &=& e^{-2\si}\,\left[{\bar R} - 6({\bar \na}\si)^2
- 6 {\bar \Box}\si \right]\,,
\label{R}
\\
R_{\mu\nu} &=& {\bar R}_{\mu\nu}
- 2 ({\bar \na}_\mu {\bar \na}_\nu \si)
- {\bar g}_{\mu\nu}({\bar \Box}\si)
+ 2 ({\bar \na}_\mu \si) ({\bar \na}_\nu \si)
- 2 {\bar g}_{\mu\nu} ({\bar \na}\si)^2\,,
\label{Ricci}
\\
R_{\mu\nu\al\be} &=& e^{2\si}\,\left[
{\bar R}_{\mu\nu\al\be}
+ \left( {\bar g}_{\mu\be}{\bar g}_{\al\nu}
- {\bar g}_{\mu\al}{\bar g}_{\be\nu} \right)
({\bar \na}\si)^2
\right.
\nonumber
\\
&+& \left.
  \left( {\bar g}_{\al\nu}{\bar \na}_\mu{\bar \na}_\be\si
- {\bar g}_{\al\mu}{\bar \na}_\nu{\bar \na}_\be\si
+ {\bar g}_{\be\mu}{\bar \na}_\nu{\bar \na}_\al\si
- {\bar g}_{\be\nu}{\bar \na}_\mu{\bar \na}_\al\si \right)
\right.
\nonumber
\\
&+&
\left.
   \left( {\bar g}_{\al\mu}{\bar \na}_\nu \si {\bar \na}_\be\si
- {\bar g}_{\al\nu}{\bar \na}_\mu \si {\bar \na}_\be\si
- {\bar g}_{\be\mu}{\bar \na}_\nu \si {\bar \na}_\al\si
+ {\bar g}_{\be\nu}{\bar \na}_\mu \si {\bar \na}_\al\si\right)
\right]\,.
\label{Riemann}
\eeq
For the metric of interest we arrive at the relations
for the non-zero components of the above curvatures
\beq
R &=& e^{-2\si}\,\left[{\bar R} - 6{\si^\prime}^2
- 6 \si^{\prime\prime} \right]\,,
\label{R-sigma}
\\
R_{\eta\eta} &=& - 3 \si^{\prime\prime}
\,,\qquad
R_{ij} = {\bar g}_{ij} \left(\frac13\,{\bar R}
- \si^{\prime\prime} - 2{\si^\prime}^2\right)\,,
\label{Ricci-sigma}
\\
R_{\eta i \eta k} &=& - e^{2\si}\,\si^{\prime\prime}\,{\bar g}_{ij}
\,,\qquad
R_{ij\,kl} = e^{2\si}\,\left(\frac16\,{\bar R}
- {\si^\prime}^2 \right)\,
\left( {\bar g}_{ik}{\bar g}_{jl}
- {\bar g}_{il}{\bar g}_{jk} \right)\,,
\label{Riemann-sigma}
\eeq
where ${\bar R}=-6k=const$.

Using the relations (\ref{R-sigma}) - (\ref{Riemann-sigma}), after
some algebra we can rewrite the elements of the action (\ref{3})
in the form \beq \sqrt{-g}R^3 &=& e^{2\si}\,\left[{\bar R} -
6{\si^\prime}^2 - 6 \si^{\prime\prime} \right]^3\,, \label{x1}
\\
\nonumber
\\
\sqrt{-g}RR_{\mu\nu}R^{\mu\nu} &=& e^{2\si}\,
\left[\frac13\,{\bar R}^3
- 2{\bar R}^2 \left(2\si^{\prime\prime} + 3{\si^\prime}^2\right)
+ 12{\bar R} \left(2{\si^{\prime\prime}}^2
+ 4{\si^{\prime\prime}}{\si^\prime}^2
+ 3{\si^\prime}^4 \right)
\right.
\nonumber
\\
&-&
\left.
 72\left({\si^{\prime\prime}}^3
+ 2{\si^{\prime\prime}}^2{\si^\prime}^2
+ 2{\si^{\prime\prime}}{\si^\prime}^4
+ {\si^\prime}^6 \right)
\right]\,,
\label{x2}
\\
\nonumber
\\
\sqrt{-g}RR_{\mu\nu\al\be}R^{\mu\nu\al\be} &=& e^{2\si}\,
\left[\frac13\,{\bar R}^3
- 2{\bar R}^2 \left(\si^{\prime\prime} + 3{\si^\prime}^2\right)
+ 12{\bar R} \left(2{\si^{\prime\prime}}^2
+ 2{\si^{\prime\prime}}{\si^\prime}^2
+ 3{\si^\prime}^4 \right)
\right.
\nonumber
\\
&-&
\left.
 72\left({\si^{\prime\prime}}^3
+ {\si^{\prime\prime}}^2{\si^\prime}^2
+ {\si^{\prime\prime}}{\si^\prime}^4
+ {\si^\prime}^6 \right)
\right]\,,
\label{x3}
\\
\nonumber
\\
\sqrt{-g}R_{\mu\nu}R^{\nu\al}R_\al^\mu
&=& e^{2\si}\,
\left[\frac19\,{\bar R}^3
- {\bar R}^2 \left(\si^{\prime\prime} + 2{\si^\prime}^2\right)
+ 3{\bar R} \left(\si^{\prime\prime} + 2{\si^\prime}^2\right)^2
\right.
\nonumber
\\
&-&
\left.
6\left(5{\si^{\prime\prime}}^3
+3{\si^{\prime\prime}}^2{\si^\prime}^2
+6{\si^{\prime\prime}}{\si^\prime}^4
+4{\si^\prime}^6\right)
\right]\,,
\label{x4}
\\
\nonumber
\\
\sqrt{-g}R_{\mu\nu\al\be}R^{\al\be\rho\la}R^{\mu\nu}\,_{\rho\la}
&=& e^{2\si}\,
\left[
\frac19\,{\bar R}^3
- 2{\bar R}^2 {\si^\prime}^2
+ 12{\bar R} {\si^\prime}^4
- 24\left({\si^{\prime\prime}}^3 +{\si^\prime}^6 \right)
\right]\,,
\label{x5}
\eeq

The sufficient condition of the reduction to a metric-scalar
theory can be easily found by using analogy with the
$\sqrt{-g}R^3$ case considered in the previous sections. In the
general case this condition has the form \beq \sqrt{-g}\, \left(
x_1R^3 + x_2RR_{\mu\nu}^2 + x_3RR_{\mu\nu\al\be}^2 + x_4
R_{\mu\nu}^3 + x_5 R_{\mu\nu\al\be}^3 \right) = e^{2\si}\,
\left[y_1{\bar R}^3 + y_2 \si^{\prime\prime} +
y_3{\si^\prime}^2\right]^3\,, \label{condition} \eeq where
$y_{1,2,3}$ are some additional arbitrary coefficients. Thus we
obtain 10 algebraic equations for the 8 variables $x_{1,2,3,4,5}$
and $y_{1,2,3}$. In fact, the number of the variables can be
immediately reduced to 7 by noticing that the coefficient $x_5$
can not be made zero in the string induced gravity by means of the
metric reparametrization, while all other coefficients can
\cite{zwei}. Therefore without losing generality we can set
$x_5=1$. Now, as far as the number of equations is much greater
than the number of independent variables, it is not certain that
the solution of these equations exist. As we shall immediately
see, it exist only for some particular, but the most relevant
case. Let us remember that our prime interest is the inflationary
epoch, where the space curvature is negligible. Then, as a first
step we can look for the solution in the simplest $k=0$ case,
where ${\bar R}=0$ and the number of equations is even smaller
than the number of independent variables. In this case the
equations become \beq - y_2^3 &=& 216x_1 + 72x_2 + 72x_3 + 30x_4
+24\,, \label{e1}
\\
- 3y_2^2y_3 &=& 648x_1 + 144x_2 + 72x_3 + 18x_4\,,
\label{e2}
\\
- 3y_2y_3^2 &=& 648x_1 + 144x_2 + 72x_3 + 36x_4\,,
\label{e3}
\\
- y_3^3 &=& 216x_1 + 72x_2 + 72x_3 + 24x_4 - 24\,.
\label{e4}
\eeq

Using the pairs of equations (\ref{e2}), (\ref{e3}) and
(\ref{e1}), (\ref{e4}) we obtain the relation \beq
y_3y_2(y_3-y_2)=(y_3-y_2)(y_3^2+y_3y_2+y_2^2)=6x_4\,. \label{e5}
\eeq These equations can be satisfied only for $y_2=y_3=y$ and
$x_4=0$. Let us remark that the constraint $y_1=y_2$ means that
possible dependence on the conformal factor for the theories
reducible to metric-scalar models can be only the same as for the
$\sqrt{-g}R^3$-case. All other choices are not reducible. The
general solution corresponds to 
\beq
x_1=-\frac{x}{3}-\frac29+\frac{y^3}{216} \,,\qquad
x_2=-2x-1+\frac{y^3}{36} \,,\qquad x_3=x \,,\qquad x_5=1\,,
\label{e6} 
\eeq 
where $x,y$ are arbitrary parameters. It is easy
to see that if the constrains (\ref{e6}) are satisfied, we have
the same reduction to a metric-scalar model as in the case of the
higher derivative term 
\beq 
S \,=\,-\,\frac{y^3}{216}\,\int d^4x
\sqrt{-g}\,R^3\,. 
\label{equivalent} 
\eeq 
At the level of the
metric-scalar model we have, in the string-induced case, exactly
the same situation as in the case of the term (\ref{equivalent}).

For the sake of completeness, we consider the more complicated
$k\neq 0$ case. An important observation is that the constraints
(\ref{e6}) must hold also for $k \neq 0$. Then elementary analysis
shows that in this case there are no solutions of eq.
(\ref{condition}). Therefore, mathematically the reduction to the
metric-scalar model can not be exact in the general case, but only
an approximate one. However the quality of this approximation is
indeed excellent because the role of $k$ during inflation is
negligible.

\section{Metric dual for the Eddington-like gravity}

As another illustration of the effectiveness of our approach,
consider the Eddington-like action of gravity \cite{eddrev,schro}
\footnote{Recently, the Eddington-like action (\ref{ed}) has
became an object of a stronger interest (see, e.g.
\cite{noteonbigravity,max} and references therein).}, 
\beq
S_{\mbox{\tiny Edd}} 
= \alpha \int \sqrt{|R_{\mu \nu}|} \, d^4 x. 
\label{ed} 
\eeq 
Here
and in what follows we use the notation \ $|R_{\mu \nu}|=|\det
(R_{\mu \nu})|$, furthermore $\alpha$ is a dimensionless parameter
and $R_{\mu \nu}$ is the (symmetric) Ricci tensor constructed from
the symmetric affine connection $\Gamma_{\mu \nu}^\lambda$. The
action (\ref{ed}) is equivalent to the vacuum Einstein-Hilbert
action with a non-zero cosmological constant
\cite{edddual,noteonbigravity}. In what follows we will show it is
straightforward to achieve the same result by using the method we
have considered in the previous sections. Moreover, in the known
approaches \cite{edddual,FrTs-dual}, it is common to consider the
first order formalism for gravity, taking $\Gamma_{\mu
\nu}^\lambda$ to be independent from the metric $g_{\mu \nu}$. Our
method does not require this restriction and, moreover, $R_{\mu
\nu}$ can be traded for any other symmetric tensor, e.g. to some
combination of the torsion fields (e.g., the one considered in
\cite{torsi}). Of course, the corresponding model will not be
equivalent to the GR, but it can be mapped to a dual theory in the
same way as we will describe below.

The dual equivalent action should have the form 
\beq 
S_{eq} = \int
d^4x \,\big\{ J^{\mu\nu}\cdot R_{\mu\nu} - V(J^{\mu\nu}) \big\}\,,
\label{eq} 
\eeq 
where $J^{\mu\nu}$ is an auxiliary field. One can
immediately note that there is a unique functional form of the
potential function $V(J)=V(J^{\mu\nu})$ which is compatible with
the covariance of the action (\ref{eq}). Indeed, the covariance
requires that $J^{\mu\nu}$ should be a tensor density and there
must be such symmetric tensor quantity $\Phi_{\mu\nu}$, such that
\beq 
J^{\mu\nu}=\sqrt{\Phi}\Phi^{\mu\nu}\,, 
\label{J-Phi-J} 
\eeq
where \ $\Phi^{\mu\al}\cdot\Phi_{\al\nu}=\de^\mu_\nu$ \ and \
$\Phi=|\det (\Phi_{\mu \nu})|$. This relation (\ref{J-Phi-J}) can
be easily inverted, so we get 
\beq 
\Phi=\det (\Phi_{\mu \nu}) =
\frac{1}{\det (J^{\mu \nu})} \quad \mbox{and} \quad \Phi^{\mu\nu}
= \frac{1}{\sqrt{|\det (J^{\mu \nu})|}}\,J^{\mu\nu}\,.
\label{Phi-J} 
\eeq 
As a result, we arrive at \ $V(J^{\mu\nu}) = k
\cdot \sqrt{\Phi}$, where $k$ is some constant. It is easy to see
from the relations (\ref{Phi-J}) that this means 
\beq
V(J^{\mu\nu}) = \frac{k}{\sqrt{|\det (J^{\mu \nu}|}}\,.
\label{JPhi} 
\eeq

Now let us see whether we can arrive at the same result
(\ref{eq}), (\ref{JPhi}) by using the method described in the
previous sections. By taking derivative of the function \
$f(R_{\mu \nu}) = \al \sqrt{|R_{\mu \nu}|}$ and inverting it, we
obtain 
\beq 
J^{\mu \nu} = \frac{\pa f}{\pa R_{\mu \nu}} \qquad
\mbox{and} \qquad R_{\mu \nu} = \frac{\pa V}{\pa J^{\mu \nu}}\,.
\label{t1} 
\eeq 
The equivalence of the two formulations requires
that $J^{\mu \nu} = J^{\mu \nu}(R_{\al \be})$, \ $R_{\mu \nu} =
R_{\mu \nu}(J^{\al \be})$ and, also, that 
\beq 
J^{\mu \nu}\cdot
R_{\mu \nu} - V(J^{\mu\nu}) = \al \sqrt{|R_{\mu \nu}|}\,.
\label{t2} 
\eeq 
Now, taking the partial derivatives $\pa/\pa
J^{\mu\nu}$ of the equality (\ref{t2}), after some simple algebra
we arrive at 
\beq 
J^{\mu \nu} = \frac{\al}{2}\, \sqrt{|R_{\mu
\nu}|}\, \left(R^{-1}_{..}\right)^{\mu \nu}\,, \label{t3} 
\eeq
where $\left(R^{-1}_{..}\right)^{\mu \nu}$ means the matrix
inverse to $R_{\mu \nu}$. It is easy to obtain the relation 
\beq
\left(R^{-1}_{..}\right)^{\mu \nu} = \frac{\al}{2}\,\frac{J^{\mu
\nu}}{\sqrt{|\det (J^{\mu \nu}|)}}\,, 
\label{t4} 
\eeq 
which shows
that $J^{\mu \nu}$ is a tensor density. The corresponding tensor
field is defined through the relation (\ref{J-Phi-J}). After some
small algebra we arrive at the equation 
\beq 
R_{\mu \nu} =
\frac{2}{\al}\,\Phi_{\mu \nu} = \frac{\pa V}{\pa J^{\mu \nu}}\,.
\label{t5} 
\eeq 
Finally, integrating this equation we get the
expected result 
\beq 
V(J^{\mu \nu}) = \frac{4}{\al\,\sqrt{ |\det
(J^{\mu\nu})|}} \,=\,\frac{4}{\al}\,\,\sqrt{\Phi}\,, 
\label{t6}
\eeq 
that is nothing else but Eq. (\ref{JPhi}) with $k=4/\al$.

It is clear that the consideration presented above opens
the way for some interesting applications. One can, at the
first place, identify the auxiliary tensor quantity
$\Phi_{\mu\nu}$
directly with the space-time metric $g_{\mu\nu}$, but this
is not the only one possible choice. Let us note that the
real identification of the metric occurs when matter is
introduced into the theory. One can assume, for instance,
that there is a scaling relation
$\Phi_{\mu\nu}=\la \cdot g_{\mu\nu}$ or even more
complicated one, like $\Phi_{\mu\nu} = B(x) \cdot g_{\mu\nu}$,
where the scalar field $B(x)$ depends on the space-time
coordinates. The corresponding generalization of the
Eddington-like gravity theory looks interesting and
perhaps deserves further discussion. The method presented
here can be immediately extended to the case when
$R_{\mu\nu}$ in the initial action (\ref{ed}) is traded
for some other tensor, e.g. for $R_{\mu\nu}+Cg_{\mu\nu}$.
In this case the auxiliary field $\Phi_{\mu\nu}$ is
naturally identified with the second metric and we arrive
at the bimetric theory of gravity \cite{noteonbigravity}.

\section{A brief note on quantum (non)equivalence}

In the previous sections 2-4 and 5 we have presented several 
examples of classically equivalent theories. It looks interesting 
to see what happens with this equivalence at the quantum level. 
It is well known that the quantum equivalence does not imply 
the quantum one. One can find statements of this type in Refs. 
\cite{H.J.Sch-94,Deru1} and also in \cite{SalHin}. Of course,
classical equivalence leads to quantum equivalence at the level
of tree diagrams or imaginary parts of one-loop diagrams which, 
in particular, describe creation of real particles and field 
perturbations by external gravitational fields and do not need 
renormalization. That is why primordial spectra of scalar and tensor 
perturbations generated during inflation in $f(R)$ gravity (\ref{1}) 
(in the models which admit it, say, in the $R+R^2$ model \cite{S80}) 
coincide with those for inflationary models with a non-minimally
coupled scalar field in the limit of a large negative coupling 
$|\xi|\gg 1$, see e.g. \cite{CY08,BKS08,LNW11}, since the latter
models reduce to the form (\ref{2}) after neglecting the kinetic
term of the scalar field that is justified in this limit. Note
that the Higgs inflation \cite{BS08} belongs to this class, too.

The problems of establishing general quantum equivalence become 
much more complicated when taking into account the need for 
renormalization. For the sake of simplicity we consider the 
theories (\ref{1}) and (\ref{2}), but the generalizations to 
other cases are indeed possible. 

The quantum equivalence at the one-loop level means, at the 
first place, that the relations like $\psi\,=\,f^\prime(R)$ 
and $R\,=\,V^\prime(\psi)$ do hold for the one-loop 
counterterms. In reality, this is not necessary so, because 
these relation may also require renormalization. Moreover, 
even if the one-loop divergences do satisfy this requirement, 
it is very unlikely that some sort of relation between the 
two (classically equivalent) theories will hold beyond one-loop
approximation. 

The main difficulty of discussing the equivalence of the 
two formulations, e.g., (\ref{1}) and (\ref{2}), or 
(\ref{s2-ini}) and (\ref{s1}) -- is the fact that the 
corresponding theories are not renormalizable. At the same 
time, even one-loop 
divergences may be non-equivalent, as it happens with the 
tensor-scalar model in different conformal frames 
\cite{spec,Kam-11}. Indeed, the non-equivalence here 
means that one can not easily find an explicit transformation 
which would link the two expressions for one-loop counterterms 
in the two representations. At the same time, since both 
theories are non-renormalizable, one can speak about certain 
qualitative equivalence in a sense that, in both frames, the 
necessary counterterms have the structure distinct from the 
one of the initial action. 

One can  look from another side and compare the UV completion 
of the two, classically equivalent, theories. It is easier 
to perform such discussion for a more general cases of the 
theories (\ref{s2-ini}) and (\ref{s1}). The theory (\ref{s1})
with ``frozen'' scalars $\psi^i$ is renormalizable 
\cite{stelle} (see also \cite{book} for a more detailed 
introduction). Therefore, in order to construct renormalizable 
theory out of (\ref{s1}), one has to complete the action by 
certain second- and fourth-derivative terms constructed from 
the scalars $\psi^i$. In case of one scalar field this 
completion has been considered in \cite{eli} and more recently 
in \cite{Weinberg-2008} in relation to inflation. The
generalization to the many-scalar case is obvious, but 
the output would be quite cumbersome. The situation with the 
general higher-derivative model (\ref{s2-ini}) is much more
complicated. First of all, there is a very strong difference  
between polynomial and non-polynomial functions $f(X_i)$. 
In the last case the problem of quantum formulation is 
unclear (despite potentially interesting, see e.g., 
\cite{Tomb}). Contrary to that, in the polynomial case the 
prescription for constructing renormalizable and 
super-renormalizable theories of quantum gravity is known 
\cite{highderi}. If the highest power of curvature tensor 
in $f(X_i)$ is $N \geq 3$, one has to introduce into Lagrangian 
all possible covariant term of this dimension, including 
the additional terms of the following form:
\beq
\sum_{k=0}^{N-2} 
\big(
\al_k R_{\mu\nu}\Box^{k}R^{\mu\nu} + \be_k R\Box^{k}R
\big)\,. 
\label{higder}
\eeq 
If the largest order coefficients $\al_{N-2}$ and  
$\be_{N-2}$ are nonzero, the theory is 
super-renormalizable \footnote{The renormalizability by 
power counting in this theory is exactly the same as in 
the Horava-Lifshits gravity \cite{Horava}.}. 

Now we are in a position to compare the renormalizability 
properties of the theories (\ref{s2-ini}) and (\ref{s1}). 
The UV completions described above correspond to the 
possible counterterms and, in these two cases, these 
counterterms are dramatically different. For example,
the UV completion in the case of scalar-tensor theory   
(\ref{s1}) describes the propagation of only two spin-2 
states, namely of graviton and of the massive spin-2 ghost. 
At the same time the UV completion of the theory 
(\ref{s2-ini}) has (for $N\geq 3$) at least one more 
spin-2 massive particle \cite{highderi}. Thus, we can 
conclude that the renormalization properties of the 
two classically equivalent non-renormalizable theories 
are, in general, quite different. In particular, they 
have very distinct UV completions and, consequently, 
very distinct structure of couterterms. 

\section{Conclusion}

We have described in the previous sections how to perform a
mapping of a gravitational theory of the type $f(X^i)$ into a
theory with auxiliary scalar fields. The number of these fields is
determined by the rank of the Hessian matrix of $f(X^i)$. The
scheme which we have described is more general than the ones known
before. In particular, it enables one to deal with the constrained
case and leads to an auxiliary-field representations for the
actions which look like renormalization-group corrected vacuum
actions in gravity theories like Eq. (\ref{cov-EB}). On the top of
this we have formulated the general conditions for the exponential
expansion of the Universe and extended the analysis for the
string-inspired case, where the treatment with auxiliary scalars
is possible only for a FRW-like solution. Finally, we have shown
that our approach is perfectly applicable to the Eddington-like
gravity models and to a wide class of their extensions. Depending
on the initial model, our approach shows either how the metric
emerges in a theory which had, initially, only affine connection,
or leads to a bimetric theories of gravity.

\acknowledgments
D.R. thanks the Departamento de Matem\'atica Aplicada at Unicamp, 
SP, Brazil, where part of this work was done, M. Ba\~nados and A. 
Gomberoff for useful comments on the Eddington action and FAPESP 
for partial financial support. F.S. and I.Sh. are grateful to 
FAPEMIG and CNPq for partial support. I.Sh. was also supported by 
ICTP visiting program. A.S. acknowledges the
RESCEU hospitality as a visiting professor. He was also partially
supported by the Russian Foundation for Basic Research under grant
09-02-12417-ofi-m.
\vskip 10mm

\renewcommand{\baselinestretch}{0.9}

\end{document}